\documentclass[a4paper,10pt,twoside]{cpc-hepnp}
\usepackage{multicol}
\usepackage{graphicx}
\usepackage{booktabs}
\usepackage{amssymb,bm,mathrsfs,bbm,amscd}
\usepackage[tbtags]{amsmath}
\usepackage{lastpage}
\usepackage{graphicx}
\usepackage{caption}
\usepackage{subcaption}
\usepackage{physics}
\usepackage{comment}
\usepackage{xcolor}

\begin{document}


\fancyhead[c]{\small Chinese Physics C~~~Vol. xx, No. x (2017) xxxxxx}
\fancyfoot[C]{\small 010201-\thepage}

\footnotetext[0]{Received Day Month Year}

\title{Antiproton identification below threshold with AMS-02 RICH detector\thanks{Supported by the  China Scholarship Council (CSC) under Grant No.201306380027.}}

\author{%
      Zi-Yuan Li$^{1,2;1)}$\email{ziyuan.li@cern.ch}%
\quad Carlos Jose Delgado Mendez$^{3)}$%
\quad Francesca Giovacchini$^{3)}$%
\quad Sadakazu Haino$^{2)}$
\quad Julia Hoffman$^{4)}$}%
\maketitle

\address{%
$^1$ School of Physics, Sun Yat-Sen University, Guangdong, 510275\\
$^2$ Institute of Physics, Academia Sinica, Taipei, 11529\\
$^3$ Centro de Investigaciones Energeticas, Mediambientales y Tecnologicas (CIEMAT), Madrid, E-28040\\
$^4$ Physics and Astronomy Department, University of Hawaii, Hawaii, 96822\\
}

\begin{abstract}
The Alpha Magnetic Spectrometer (AMS-02) was installed on the International Space Station (ISS) and it has been collecting data successfully since May 2011. The main goals of AMS-02 are the search for cosmic anti-matter, dark matter and the precise measurement of the relative abundance of elements and isotopes in galactic cosmic rays. In order to identify particle properties, AMS-02 includes several specialized sub-detectors. Among them, the AMS-02 Ring Imaging Cherenkov detector (RICH) is designed to provide a very precise measurement of the velocity and electric charge of particles. We describe a method to reject the dominant electron background in antiproton identification with the use of the AMS-02 RICH detector as a veto for rigidities below 3 GV. Ray tracing integration method is used to maximize the statistics of $\bar{p}$ with the lowest possible $e^{-}$ background, providing 4 times rejection power gain for $e^{-}$ background with respect to only 3\% of $\bar{p}$ signal efficiency loss. By using the collected cosmic-rays data, $e^{-}$ contamination can be well suppressed within 3\% with $\beta \approx 1$, while keeping 76\% efficiency for $\bar{p}$ below the threshold.
\end{abstract}



\footnotetext[0]{\hspace*{-3mm}\raisebox{0.3ex}{$\scriptstyle\copyright$}2016
Chinese Physical Society and the Institute of High Energy Physics
of the Chinese Academy of Sciences and the Institute
of Modern Physics of the Chinese Academy of Sciences and IOP Publishing Ltd}%

\begin{multicols}{2}

\section{Introduction}

\begin{center}
\includegraphics[width=0.38\textwidth]{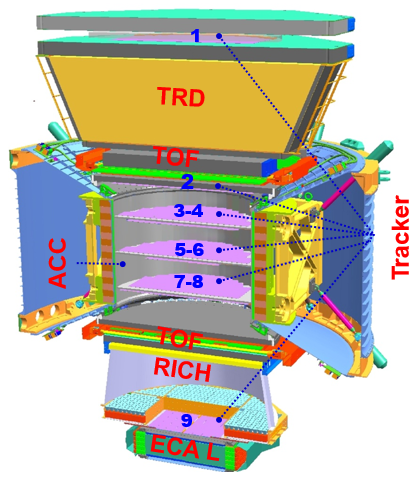}
\figcaption{\label{ams}Schematic view of AMS-02 detector. }
\end{center}

The AMS-02 detector is a general purpose, high energy physics detector described in detail in ~\cite{lab1}. It is designed to measure cosmic ray particles in the rigidity (defined as R=p/Z in natural units, where R is rigidity, p is particle momentum and Z its charge) range from 1 GV to a few TV. The detector consists of the following subsystems: a nine-layer silicon tracker~\cite{labtrk}, seven of which are surrounded by a permanent magnet~\cite{labmagnet}; a transition radiation detector (TRD)~\cite{labtrd}; four planes of time of flight scintillator counters (TOF)~\cite{labtof}; an array of anticoincidence counters (ACC)~\cite{labacc}; a ring imaging Cherenkov detector (RICH)~\cite{labrich}; and an electromagnetic calorimeter (ECAL)~\cite{labecal}. 

The silicon tracker, together with the magnet provides precise measurement of particle rigidity, its incoming direction and charge. The maximum detectable rigidity (MDR) over the 3 m lever arm from layer 1 (L1) to layer 9 (L9) is $\sim$2 TV for electrons, 2 TV for protons and 3.2 TV for helium. The TRD is designed to distinguish between hadrons and leptons with the use of transition radiation process. The four planes of TOF scintillator counters (two above and two below the magnet) provide fast trigger, particle velocity, incoming direction and charge. The 16 ACC counters form a cylindrical shell between the magnet and the tracker. Adjacent counters, combined, provide 8 readout sectors to reject cosmic rays entering the tracker from the side. RICH detector measures particle velocity and charge. Combined with the tracker information, the RICH and the TOF allow for particle mass separation. The ECAL measures energy and incidence direction of electrons, positrons and photons. Also, the ECAL allows for separation between hadrons and leptons independently of the TRD. A detailed schematic view of AMS-02 detector is presented in Fig.~\ref{ams}.

\section{Description of the RICH detector}

\begin{center}
\includegraphics[width=0.43\textwidth]{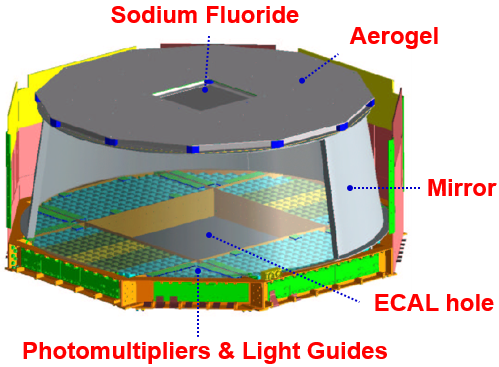}
\figcaption{\label{rich}Layout of AMS-02 RICH detector. }
\end{center}

The RICH detector of AMS-02 is located between the Lower TOF and the ECAL. By providing precise measurements of velocity and absolute charge of incoming cosmic-rays, it plays a very important role in particle identification. As illustrated in Fig.~\ref{rich}, from top to the bottom, the RICH detector is composed of three main parts: a radiator plane where Cherenkov light is emitted, a high reflectivity mirror to increase detection efficiency, and a detection plane made of photomultiplier matrix and light guides~\cite{lab2,lab3,lab4}. The detection plane has a hole at the center to minimize the amount of material in front of the ECAL. A cone of Cherenkov light is emitted when an incoming charged particle crosses the radiator material with a velocity ($\beta$) greater than the Cherenkov threshold ($\beta_{thr} = \dfrac{1}{n}$) for this medium with refractive index (n). The aperture angle of the emitted photons with respect to the radiating particle direction is known as the Cherenkov angle ($\theta_c$), which is connected to velocity according to equation Eq.~\ref{eq1}~\cite{lab5}:
\begin{eqnarray}
\label{eq1}
\beta = \dfrac{1}{n \cdot cos(\theta_{c})}.
\end{eqnarray}

In AMS-02, the RICH radiator plane includes two different radiators: 16 squared tiles of Sodium Fluoride (NaF) with dimensions $8.5\times8.5\times0.5$ cm$^3$ in the center covering $\sim 10\%$ of the RICH acceptance, and 92 silica aerogel tiles (AGL) with dimensions $11.5\times11.5\times2.5$ cm$^3$ surrounds the NaF radiator. The Cherenkov threshold for AGL is $\beta_{thr,AGL} = 0.95$, while for NaF, it is $\beta_{thr,NaF} = 0.75$. As compared to AGL, the NaF radiator allows to detect particles in a wider velocity range. Moreover, particles passing through NaF will generate Cherenkov radiation with larger Cherenkov angle, which increases the detector acceptance for particle trajectories pointing towards the ECAL.

In the following sections, we focus on the discussion of antiproton identification method below threshold with AGL radiator, valid for rigidities below 3 GV.

\section{Particle Identification Method}

In AMS-02, particle identification is done by using measurement from different subsystems~\cite{labantiproton}. At high rigidity ($\abs{R} > 10$ GV) electrons and hadrons (such as antiprotons, pions, or kaons) can be identified in the TRD with the TRD estimator $\Lambda_{TRD}$~\cite{zweng} and also in ECAL with the ECAL estimator based on a boosted decision tree algorithm, which is constructed from the shower shape in the ECAL~\cite{lab6,lab7,lab8}. In AMS-02 the RICH velocity resolution for charge 1 particles has been measured using in-flight calibrated data to be $\Delta\beta/\beta \sim 10^{-3}$ in the case of AGL radiator~\cite{lab9}. Therefore at middle rigidity ($\abs{R}$ between 3 to 10 GV), velocity measured with the RICH detector $\beta_{RICH}$ together with the $\Lambda_{TRD}$ is used to separate the antiproton signal from light particles ($e^{-}$ and $\pi^{-}$, $K$) background. ECAL is not used for particles passing through AGL. For $\bar{p}$ below the threshold ($\abs{R} < 3$ GV), this paper provides an innovative approach to use AMS-02 RICH detector as a threshold-type aerogel Cherenkov detector to reject the remaining electron backgound in the identification of cosmic ray antiproton after $\Lambda_{TRD}$ selection.

\subsection{Basic Idea}

The corresponding rigidity for $p$ ($\bar{p}$) with $\beta = \beta_{thr,AGL}$ is 3 GV. Therefore below 3 GV, $p$ will not produce Cherenkov light when passing through AGL, while for electrons (for which $\beta \approx 1$ in this rigidity range) Cherenkov light emission is expected. Based on this concept, we require that the number of photoelectrons~\cite{lab2} ($N_{pe}$) detected by RICH equals to zero in order to reject electrons. The $N_{pe}$ signal produced by the charged particle in the impact point of the detection plane due to energy losses in light guide have already been subtracted.
\end{multicols}

\begin{figure*}
	\centering
	\begin{subfigure}{0.4\textwidth}
    	\centering
    	\includegraphics[width=0.83\textwidth]{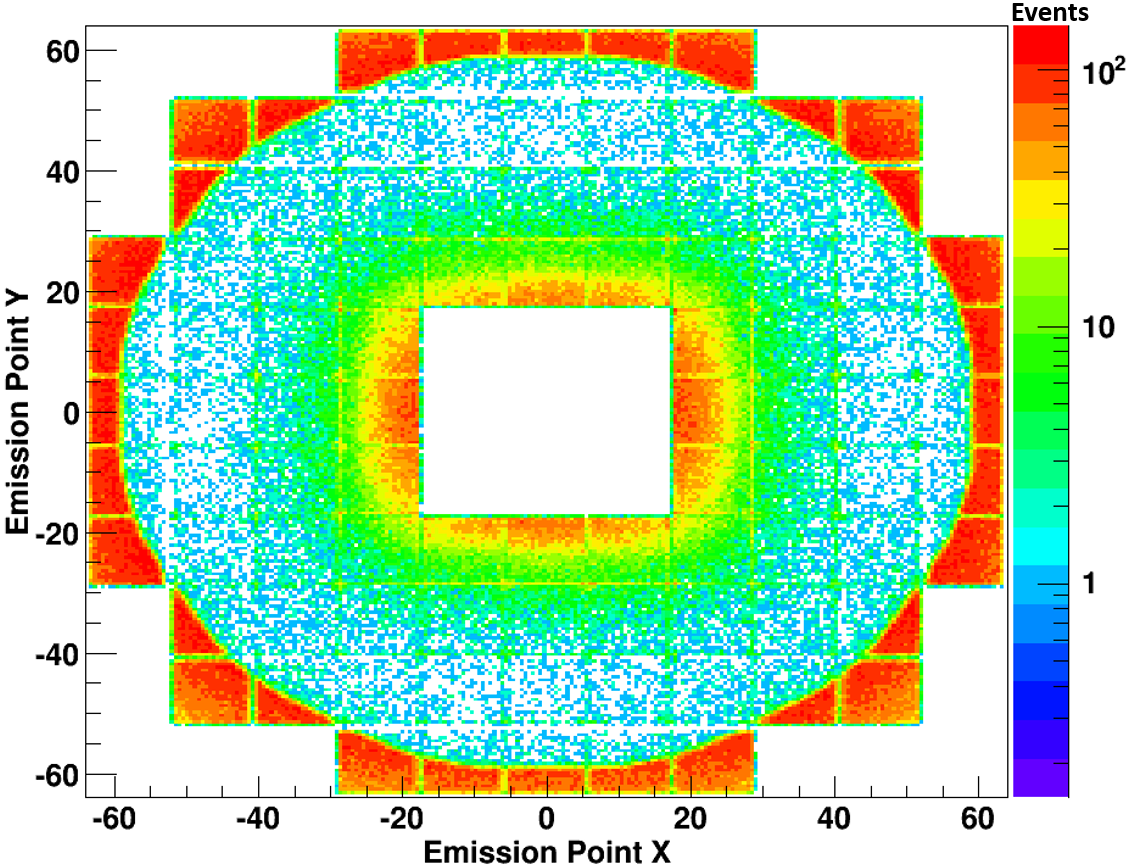}
    	\figcaption{\label{a}Simulated $e^{-}$: $N_{pe}$ = 0}
    \end{subfigure}
    \begin{subfigure}{0.4\textwidth}
    	\centering
    	\includegraphics[width=0.83\textwidth]{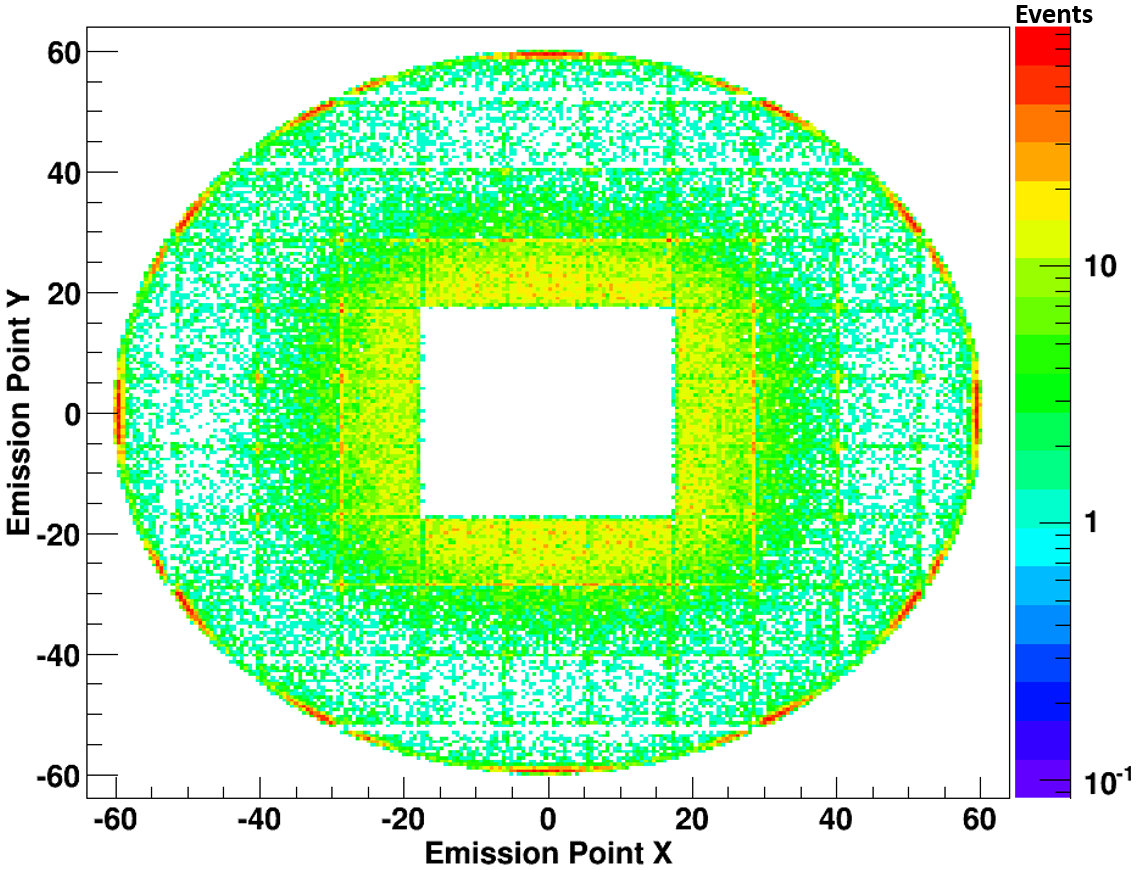}
    	\figcaption{\label{b}Simulated $e^{-}$: $N_{pe}$ = 0 \&\& $N_{Exppe} > 2$}
    \end{subfigure}
    \begin{subfigure}{0.4\textwidth}
    	\centering
    	\includegraphics[width=0.83\textwidth]{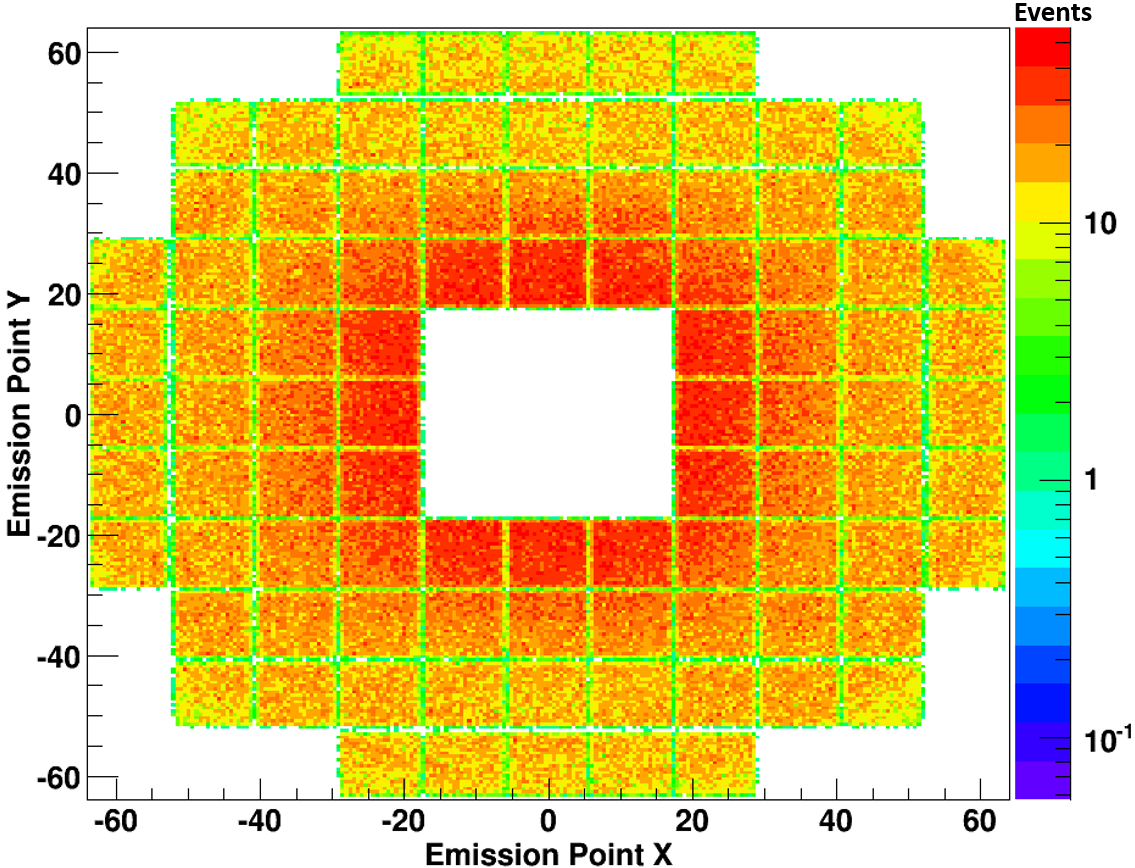}
    	\figcaption{\label{c}Simulated $p$: $N_{pe}$ = 0}
	\end{subfigure}
	\begin{subfigure}{0.4\textwidth}
		\centering
		\includegraphics[width=0.83\textwidth]{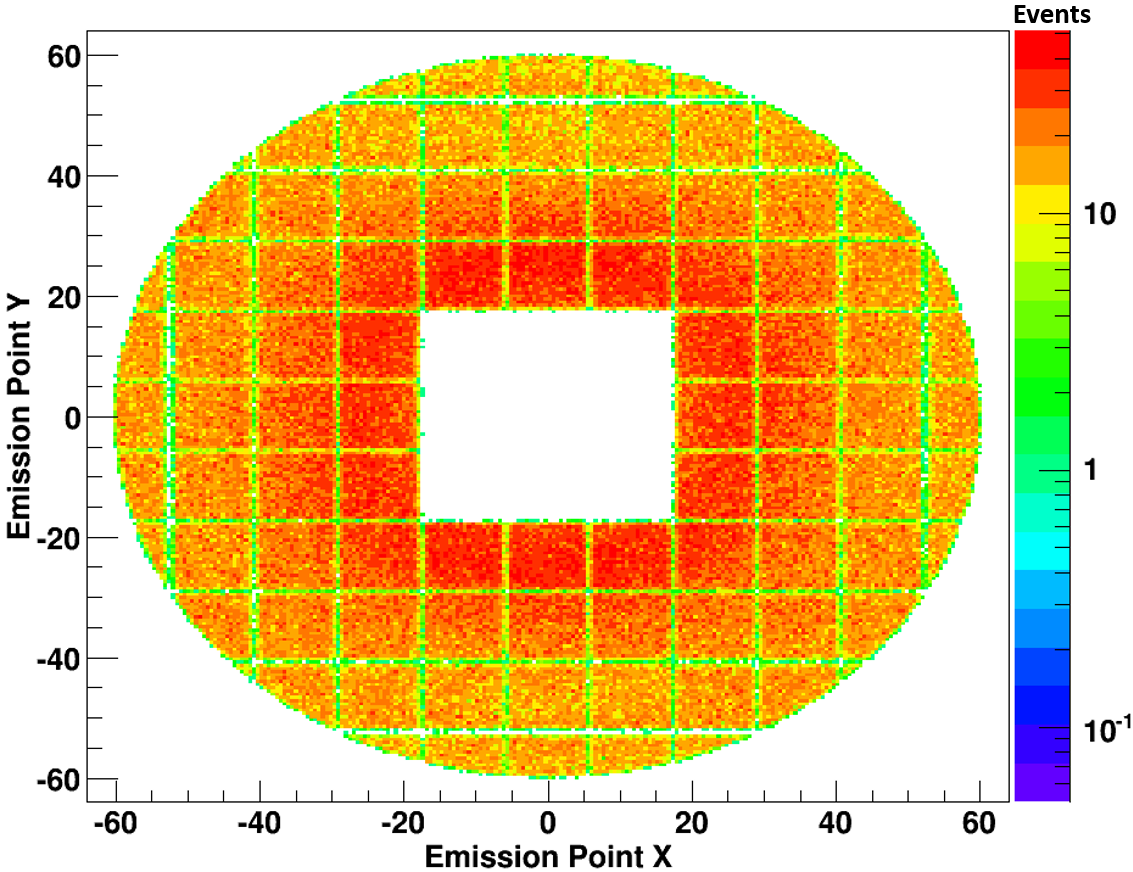}
		\figcaption{\label{d}Simulated $p$: $N_{pe}$ = 0 \&\& $N_{Exppe} > 2$}
    \end{subfigure}
    \figcaption{\label{agl}Particle emission coordinates in Aerogel in the X-Y plane of the detector.}   

\end{figure*}

\begin{multicols}{2}

However, in order to further reject electrons, additional effects should be considered. Events with zero photoelectrons could be electrons where Cherenkov radiation was lost or absorbed due to one of possible effects like Rayleigh Scattering, total reflection in aerogel radiator, reflection or absorption on the mirror surface, falling into non-active area, light guide losses, and the like. 

To see the evidence of those effects, Monte Carlo samples of electrons and protons in rigidity range 1 to 3 GV were produced in GEANT 4.10.1 package~\cite{lab10,lab11}. The samples underwent the same process of reconstruction method as cosmic ray data. From the emission coordinates of incoming particle in aerogle radiator, we can see the difference between $e^{-}$ and $p$. As shown in Fig.~\ref{a}, the majority of $e^{-}$ events after requirement of $N_{pe}=0$ tend to distribute around the region outside of the top mirror radius and near the ECAL hole region. For $p$ events (Fig.~\ref{c}), the distribution is basically uniform except the center region due to larger acceptance.

In order to maximize the statistics of $p$ ($\bar{p}$) with the lowest possible electron background given the aforementioned selection, it is necessary to perform an unbiased selection of events in the RICH detector that guarantees the number of detected photoelectrons for $e^{-}$ is greater than zero after accounting for aforementioned effects. This is done by computing the number of expected photoelectrons ($N_{Exppe}$) in an event by event basis, using a Simple Monte Carlo Approach and assuming a $\beta=1$ particle. Rejecting events with a low value of this quantity selects a geometrical region of the detector with a high probability of detection of Cherenkov photons, therefore increasing the separation power of the $N_{pe}$ selection.

\subsection{Simple Monte Carlo Approach}

For each Simple Monte Carlo Approach event, assuming it is an electron, combined with rigidity from tracker we can generate a homogeneous cone of photons with Cherenkov angle $\theta_{c}=\arccos{(\sqrt{R^2+m^2_e}/nR)}$ ($m_{e}$ is the mass of electron). Then we can use a ray tracing integration method to compute the detection efficiency after taking into account all losses and absorption. To get the number of expected photoelectrons $N_{Exppe}$, we need to multiply the detection efficiency by the expected light yield, which at $\beta = 1$ is proportional to $(n^2-1)/n^2$ and depends linearly on the sample's thickness and radiator absorption.

We require $N_{Exppe} > 2$ to select particles that pass through good geometry region and are good enough to reconstruct a ring for electron events. The cut effect can be clearly seen in Fig.~\ref{b} and Fig.~\ref{d}. For simulated $e^{-}$, the cut efficiency of $N_{Exppe} > 2$ after requiring $N_{pe} = 0$ is $\sim 20\%$ while for $p$, the efficiency is $\sim 88\%$. Using $N_{Exppe}$ from Simple Monte Carlo Approach, we can achieve the goal to maximize the statistics of $p$ ($\bar{p}$) with the lowest possible electron background.

\section{Data selection}
Five years data have been analysed starting from May of 2011 to May of 2016. During this period, over 80 billion cosmic-rays events have been recorded. Events considered in this analysis were collected during normal detector operation, i.e. during time periods when the AMS Z-axis is pointing within $40^{\circ}$ of the local zenith and when the ISS is not in the South Atlantic Anomaly.

Further selection requires a track in the inner tracker, matching a track inside the TRD. The velocity in the TOF is required to be $\beta_{TOF} > 0.3$, corresponding to a downward-going particle. Moreover, the $dE/dx$ measurements in the TRD, the TOF, and the inner tracker must be consistent with $\abs{Z}$ = 1. The accuracy of track reconstruction, the $\chi^2/d.f.$ is required to be less than 10 both in the bending and non-bending projections in order to maximize the accuracy of the track reconstruction. The track, extrapolated from the Tracker outwards should pass through both surfaces (upper and lower) of the aerogel. Good geometry region is chosen with requirement of $N_{Exppe} > 2$. The events in focus are those with rigidity below aerogel threshold for protons ($\abs{R} < 3$ GV).

Events satisfying the selection criteria are classified into two categories - positive and negative rigidity events. In positive rigidity population, $99.9\%$ are pure protons with almost no background. Among negative rigidity events there are antiprotons and several background sources : electrons, light negative mesons ($\pi^{-}$ and negligible amount of $K^{-}$), produced in the interactions of primary cosmic rays with the detector material.

To evaluate the performance of this antiproton identification method, TRD estimator $\Lambda_{TRD}$ is used as an independent detector to select pure proton and pure electron samples.

\section{Performance}
The Normalized $N_{pe}$ distribution of cosmic ray protons (in blue) and electrons (in red) in the rigidity range $1.51 - 1.71$ GV is shown in Fig.~\ref{ISSNpe}. As it can be seen from the plot, majority of protons concentrate at 0 while for electrons the distribution is wider. From a detailed analysis, the upper tail of the histogram produced by below-threshold particles can be attributed to several different sources as $\delta$-rays, accidental particle crossing the radiator, scintillation light produced in the radiator, and the like. The peak at zero for electrons is due mainly to events that cannot be reconstructed due to a statistical fluctuation of the real number of photoelectrons in the event.

\begin{center}
\includegraphics[width=0.5\textwidth]{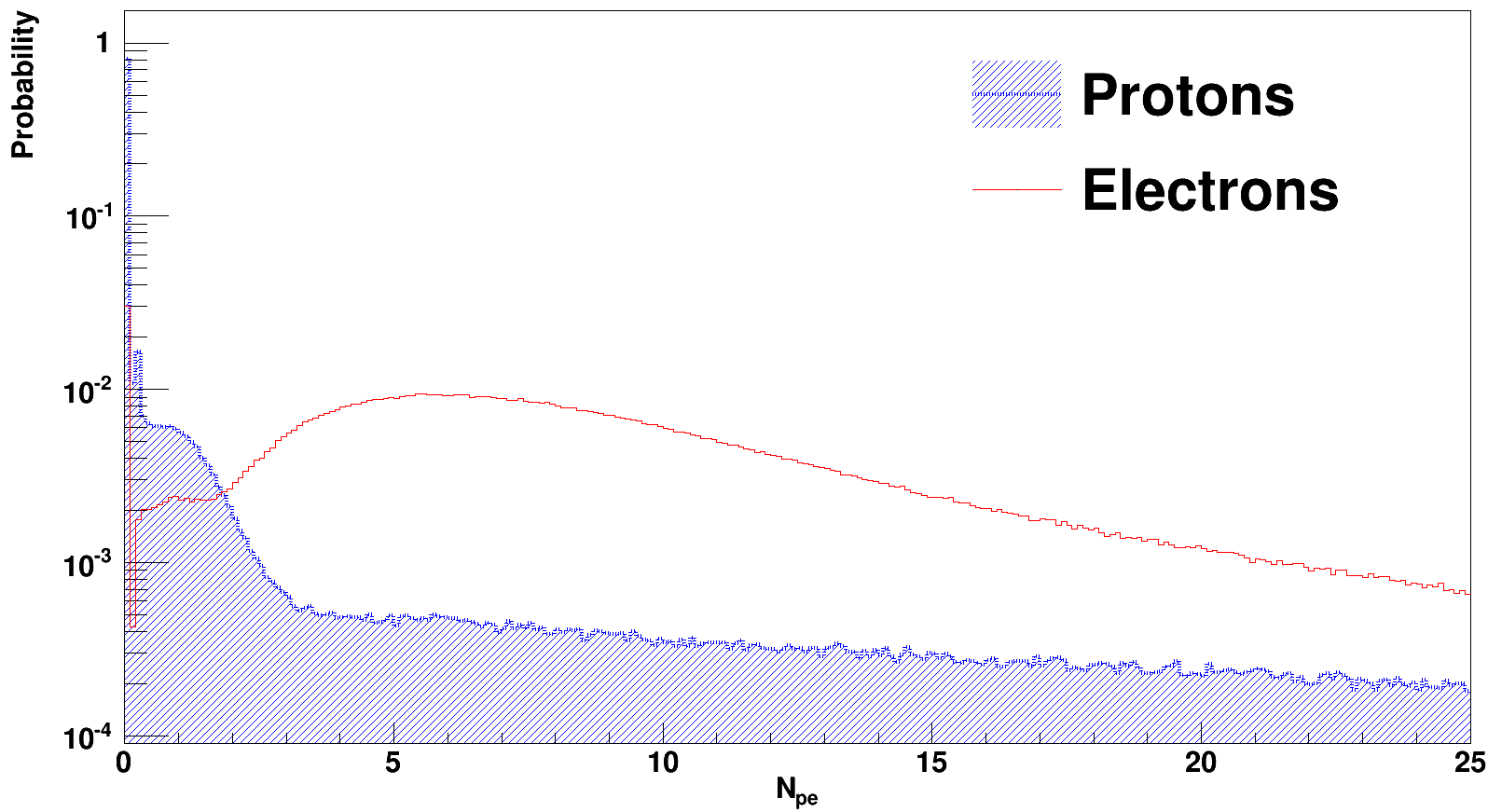}
\figcaption{\label{ISSNpe}Normalized $N_{pe}$ distribution for cosmic ray protons (in blue) and electrons (in red) in rigidity range $1.51 - 1.71$ GV. }
\end{center}

To study the performance of this antiproton identification method, we can make a tight cut by applying $N_{pe} = 0$ (together with $N_{Exppe} > 2$ cut) and see how the proton (in blue) and electron (in red) cut efficiency changes with rigidity, as shown in Fig.~\ref{NpeEfficiency_issmc}. Here the inverted filled triangle points are cosmic ray data while regular empty triangle points are simulated data. The simulated data reproduce well the result from cosmic ray data. For protons, the cut efficiency is about $76\%$ and keeps almost constant up to 3 GV. For electrons, the cut efficiency is about $3\%$.

\begin{center}
\includegraphics[width=0.5\textwidth]{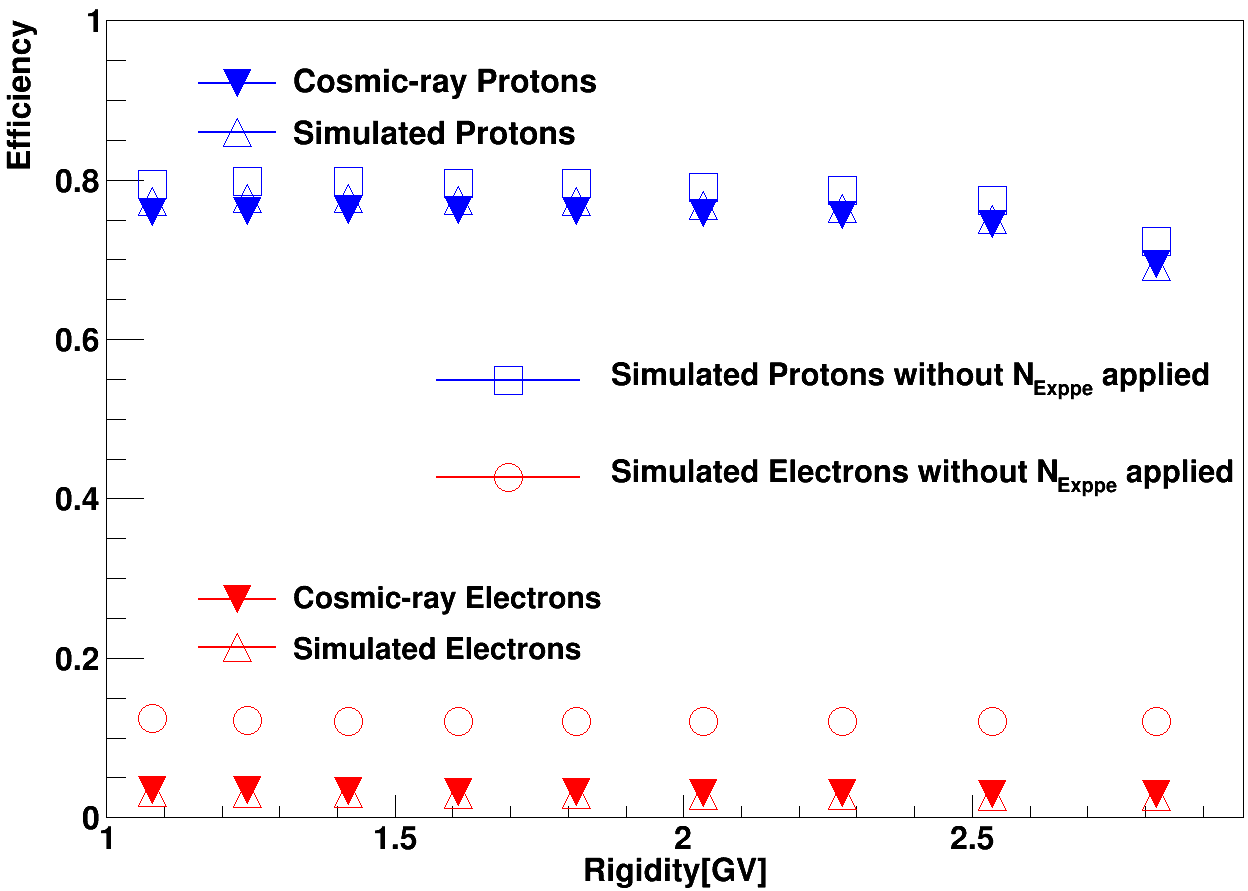}
\figcaption{\label{NpeEfficiency_issmc}$N_{pe} = 0$ cut efficiency changes with rigidity in case $N_{Exppe} > 2$ cut is applied. Inverted filled triangle points are cosmic ray protons (in blue) and electrons (in red) while regular empty triangle points are simulated data. Square points and circle points are the result of simulated protons and electrons without $N_{Exppe}$ applied.}
\end{center}

\end{multicols}

\begin{figure*}
	\centering
	\begin{subfigure}{0.45\textwidth}
    	\centering
    	\includegraphics[width=\textwidth]{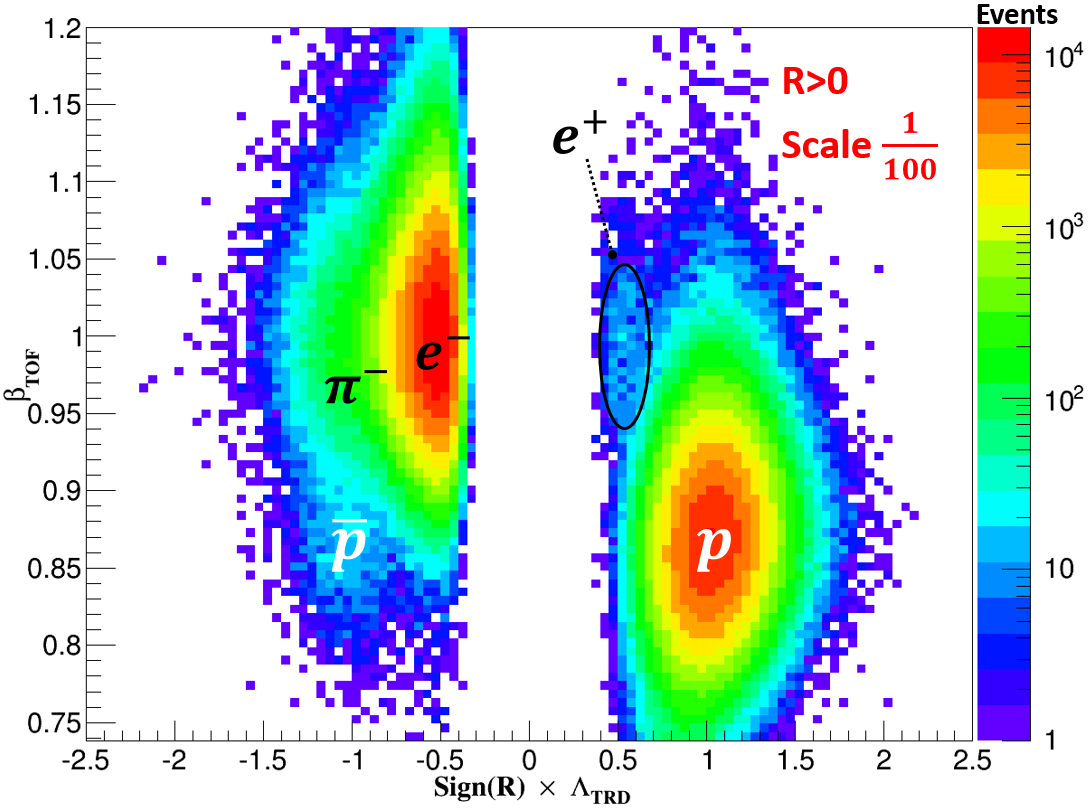}
    	\figcaption{\label{trdvstofbetaA}Before apply $N_{pe}$ = 0 \&\& $N_{Exppe} > 2$ cuts}
    \end{subfigure}
    \begin{subfigure}{0.45\textwidth}
    	\centering
    	\includegraphics[width=\textwidth]{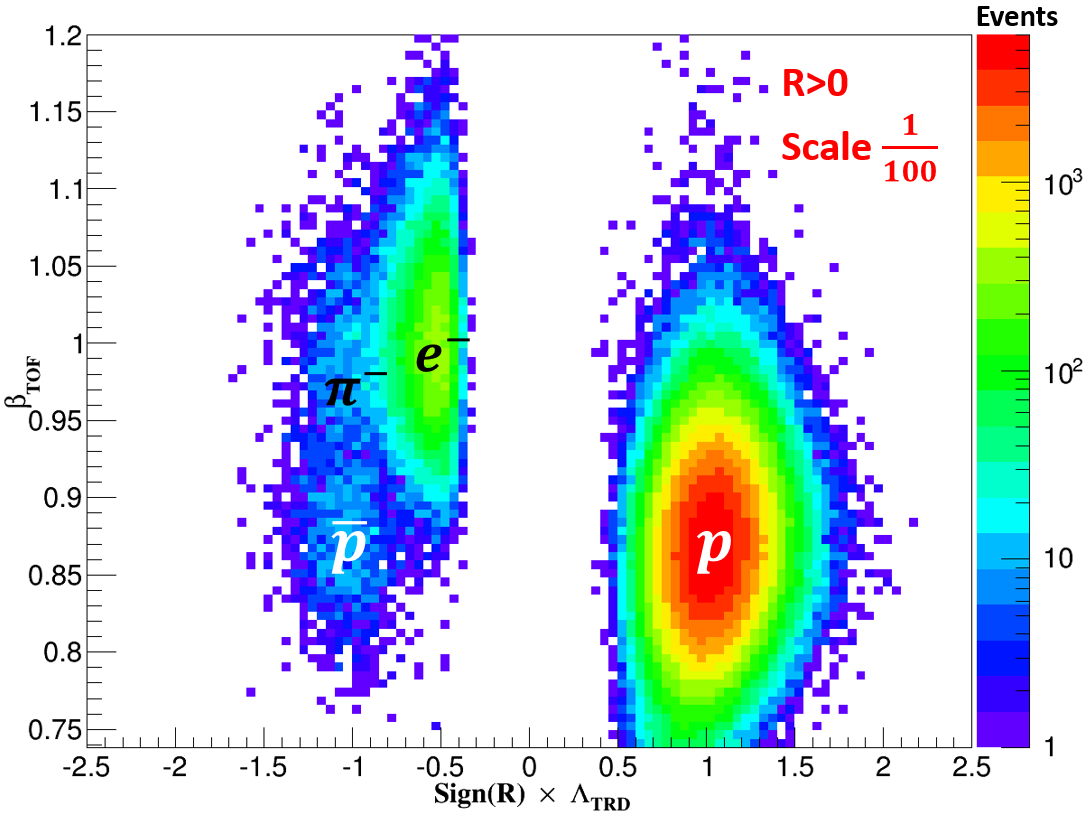}
    	\figcaption{\label{trdvstofbetaB}After apply $N_{pe}$ = 0 \&\& $N_{Exppe} > 2$ cuts}
    \end{subfigure}
    \begin{subfigure}{0.45\textwidth}
    	\centering
    	\includegraphics[width=\textwidth]{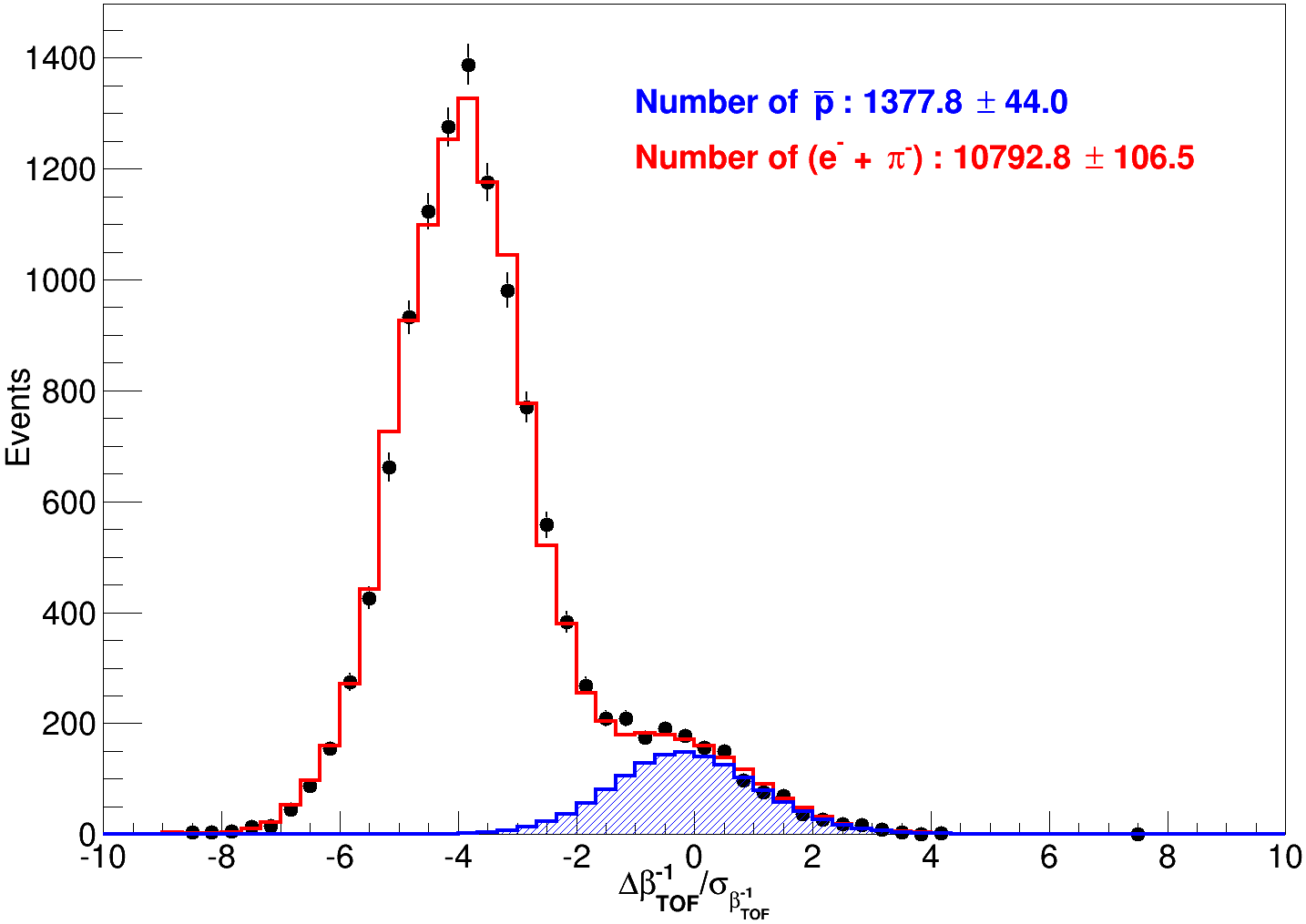}
    	\figcaption{\label{b4rv}Before apply $N_{pe}$ = 0 \&\& $N_{Exppe} > 2$ cuts}
    \end{subfigure}    
    \begin{subfigure}{0.45\textwidth}
    	\centering
    	\includegraphics[width=\textwidth]{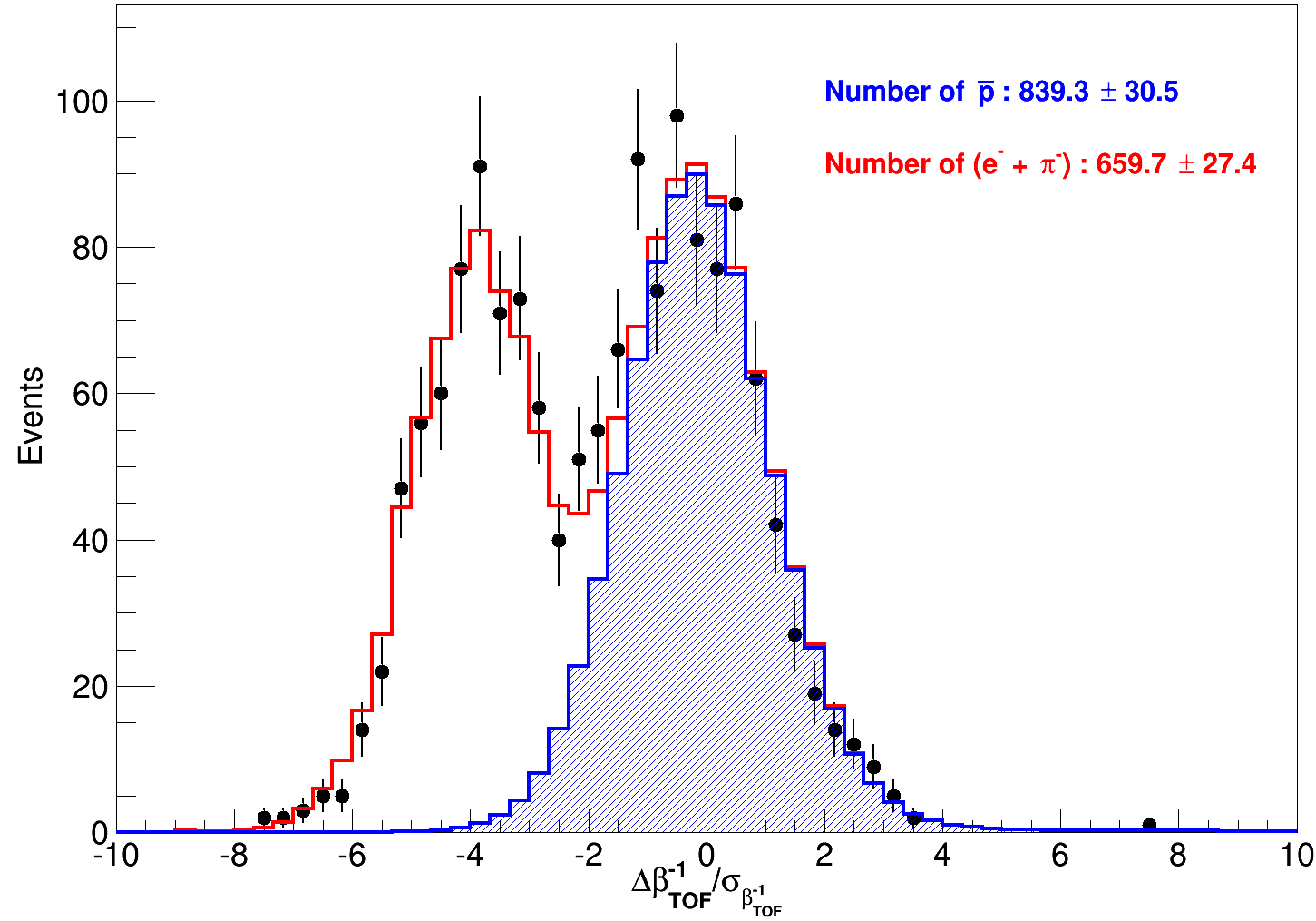}
    	\figcaption{\label{afrv}After apply $N_{pe}$ = 0 \&\& $N_{Exppe} > 2$ cuts}
    \end{subfigure}    
    \figcaption{\label{trdvstofbeta}(a) (b) : Negative rigidity and positive rigidity data samples in the ($\beta_{TOF}$ - $\Lambda_{TRD}$) plane for the absolute rigidity range $1.51 - 1.71$ GV. Scale 1/100 for positive events. (c) (d) : Distribution of measured $1/\beta_{TOF}$ with respect to the theoretical value from rigidity measurement and assumption of $p$ mass ($\Delta\beta^{-1}_{TOF}$), and normalized by the TOF $\beta^{-1}$ resolution ($\sigma_{\beta^{-1}_{TOF}}$). Black points are negative data samples after requiring $\Lambda_{TRD} > 0.8$. Template fit method is used to determine the number of $\bar{p}$ events (in blue).}
\end{figure*}

\begin{multicols}{2}

In order to demonstrate the importance of $N_{Exppe}$, $N_{pe} = 0$ cut efficiency has been studied without applying $N_{Exppe}$ cut on simulated data. As shown in Fig.~\ref{NpeEfficiency_issmc}, square points are protons (in blue) while circle points are electrons (in red). Without $N_{Exppe}$ cut, the $N_{pe} = 0$ efficiency is about 79\% for protons and about 12\% for electrons. This means that with $N_{Exppe}$ we can gain 4 times the rejection power for electrons while only 3\% efficiency of protons loss.

After apply $N_{pe} = 0$ and $N_{Exppe} > 2$ cuts on cosmic ray data, velocity measured with the TOF detector $\beta_{TOF}$ together with the $\Lambda_{TRD}$ is used to separate the antiproton signal from light particles background. As shown in Fig.~\ref{trdvstofbetaA} and Fig.~\ref{trdvstofbetaB}, a well enhanced $\bar{p}$ signal in the ($\beta_{TOF}$ - $\Lambda_{TRD}$) plane for the absolute rigidity range 1.51 - 1.71 GV can be seen. For positive rigidity events, we scale 1/100 in the plot in order to make the change in negative part clear. To determine the number of $\bar{p}$ signal events, we require $\Lambda_{TRD} > 0.8$ and project the remaining negative rigidity events to $\Delta\beta^{-1}_{TOF}/\sigma_{\beta^{-1}_{TOF}}$ plane. $\Delta\beta^{-1}_{TOF}/\sigma_{\beta^{-1}_{TOF}}$ is the distribution of measured $1/\beta_{TOF}$ with respect to the theoretical value from rigidity measurement and assumption of $p$ mass ($\Delta\beta^{-1}_{TOF}$), and normalized by the TOF $\beta^{-1}$ resolution ($\sigma_{\beta^{-1}_{TOF}}$). As shown in Fig.~\ref{b4rv} and Fig.~\ref{afrv}. The ($e^{-}$ + $\pi^{-}$) background template is defined from the cosmic ray data samples selected using information from TRD, RICH. The $\bar{p}$ signal template (in blue) can be easily derived from cosmic ray $p$ samples. Black points are data to be fitted. From the fit result, the cut efficiency of $N_{pe} = 0$ and $N_{Exppe} > 2$ for $\bar{p}$ is 61\%, while on Fig.~\ref{NpeEfficiency_issmc} the conclusion of 76\% is the cut efficiency of $N_{pe} = 0$ cut only. With the antiproton identification method introduce in this paper, we are able to control the background to the same level as antiproton signal.

\section{Conclusions}
A dedicated method to use AMS-02 RICH detector as threshold-type aerogel Cherenkov detector has been discussed. The cut efficiency of this method derived from cosmic ray data can be well reproduced by the simulated data. We report 4 times rejection power gain for $e^{-}$ background while only 3\% of $\bar{p}$ signal efficiency loss using ray tracing integration method. Electron contamination can be well suppressed within 3\% with $\beta \approx 1$, while keeping 76\% efficiency for protons below the threshold.

With this method, antiproton signal can be enhanced in a better way from background events
of electrons and light negative mesons in the ($\beta_{TOF}$ - $\Lambda_{TRD}$) plane.

\end{multicols}

\vspace{-1mm}
\centerline{\rule{80mm}{0.1pt}}
\vspace{2mm}

\begin{multicols}{2}

\end{multicols}

\clearpage
\end{document}